\documentclass[superscriptaddress,preprint,amsmath,amssymb,floats,pre]{revtex4}

\usepackage{graphicx}

\usepackage{dcolumn}

\usepackage{bm}

\usepackage{revsymb}

\usepackage[usenames]{color}

\usepackage{subfigure}

\begin{document}

\newcommand{\brm}[1]{\bm{{\rm #1}}}
\newcommand{\tens}[1]{\underline{\underline{#1}}}
\newcommand{\mm}{\overset{\leftrightarrow}{m}}
\newcommand{\xv}{\bm{{\rm x}}}
\newcommand{\Av}{\bm{{\rm A}}}
\newcommand{\uv}{\bm{{\rm u}}}
\newcommand{\vv}{\bm{{\rm v}}}
\newcommand{\piv}{\bm{{\rm \pi}}}
\newcommand{\nv}{\bm{{\rm n}}}
\newcommand{\Nv}{\bm{{\rm N}}}
\newcommand{\Fv}{\bm{{\rm F}}}
\newcommand{\Zv}{\bm{{\rm \zeta}}}
\newcommand{\ev}{\bm{{\rm e}}}
\newcommand{\dv}{\bm{{\rm d}}}
\newcommand{\bv}{\bm{{\rm b}}}
\newcommand{\lv}{{\bm{l}}}
\newcommand{\rv}{\bm{{\rm r}}}
\newcommand{\id}{\tens{\mathbb I}}
\newcommand{\bhv}{\hat{\bv}}
\newcommand{\bh}{\hat{b}}
\def\ten#1{\underline{\underline{{#1}}}}
\newcommand{\Ft}{{\tilde F}}
\newcommand{\Ftv}{\tilde{\mathbf{F}}}
\newcommand{\sigmat}{{\tilde \sigma}}
\newcommand{\sigmab}{{\overline \sigma}}
\newcommand{\ellv}{\mathbf{\ell}}
\newcommand{\qv}{\bm{{\rm q}}}
\newcommand{\pv}{\bm{{\rm p}}}
\newcommand{\tD}{\underline{D}}
\newcommand{\Tchange}[1]{{\color{red}{#1}}}
\newcommand{\Fa}{{\cal F}}
\newcommand{\Uv}{{\roarrow U}}
\newcommand{\Bv}{{\roarrow B}}
\newcommand{\Dt}{{\tensor {\cal D}}}
\newcommand{\cv}{{\mathbf c}}
\newcommand{\pt}{\tilde{p}}
\newcommand{\as}[1]{{\color{black}{#1}}}
\newcommand{\fbz}{1$^{st}$ BZ}
\newcommand{\cblue}{\color{black}}

\definecolor{noir}{rgb}{0.2,0.1,0.45}
\definecolor{noir}{rgb}{0,0,0}

\title{Topological sound in active-liquid metamaterials}
\author{Anton Souslov}
\affiliation{Lorentz Institute, Leiden University, Leiden, The Netherlands}
\author{Benjamin C. van Zuiden}
\affiliation{Lorentz Institute, Leiden University, Leiden, The Netherlands}
\author{Denis Bartolo}
\affiliation{Univ. Lyon, ENS de Lyon, Univ. Claude Bernard, CNRS, 
Laboratoire de Physique, F-69342 Lyon, France}
\author{Vincenzo Vitelli}
\affiliation{Lorentz Institute, Leiden University, Leiden, The Netherlands}

\date{\today}


\begin{abstract}
 Liquids composed of self-propelled particles have been experimentally realized using molecular, colloidal, or macroscopic 
constituents~\cite{Bausch2010,Bricard2013,Granick2016,Deseigne2010,SoodRamaswamy2014}. 
These active liquids can flow spontaneously even in the absence of an external drive ~\cite{Marchetti2013,Vicsek_review,Toner:2005wy}. 
Unlike spontaneous active flow ~\cite{Wioland2016,Bricard2015}, the propagation of density waves 
in confined active liquids is not well explored. Here, we exploit a mapping between density waves on top of a chiral flow and electrons in a synthetic gauge field ~\cite{Yang2015,Khanikaev2015} to lay out design principles for artificial structures termed topological active metamaterials. We design metamaterials that break time-reversal symmetry using lattices composed of annular channels filled with a spontaneously flowing active liquid. Such active metamaterials support topologically protected sound modes that propagate unidirectionally, without backscattering, along either sample edges or domain walls and despite overdamped particle dynamics. Our work illustrates how parity-symmetry breaking in metamaterial structure combined with microscopic irreversibility of active matter leads to novel functionalities that cannot be achieved using only passive materials.
\end{abstract}

\maketitle


We design active metamaterials with transport properties akin to those of quantum Hall fluids \cite{Hasan2010} by confining active liquids in periodic geometries that generate gapped density-wave spectra. Recent studies of topological acoustics have revealed that spectral bands characterized by topological invariants host (in their spectral gaps) robust mechanical states~\cite{Kane2014, Chen2014, Paulose2015} and sound modes that propagate unidirectionally along sample edges and 
interfaces~\cite{Prodan2009,Khanikaev2015,Yang2015,Nash2015,Susstrunk2015, Kariyado2015, Wang2015, He2016}.
However, the translation of topological-acoustic designs from macroscopic prototypes to soft materials has so far proven challenging, because overdamped particle dynamics overcome inertia and suppress the propagation of ordinary sound waves at the microscale. To address this challenge, we elucidate the relationship between emergent active flow and the spectrum of topological density waves in a confined liquid composed of self-propelled particles that have overdamped dynamics and align their velocities, i.e., a confined polar active liquid.

In order to obtain generic results, we use a continuum mechanics description of polar active flow. The analog of Navier-Stokes equations that describe a one-component fluid of self-propelled particles (with overdamped particle dynamics, see Supplementary Information [SI]) are the Toner-Tu equations~\cite{Toner1995,Toner:2005wy,Marchetti2013}, which in their simplest form read:
\begin{align}
&\partial_t \varrho +  v_0\bm\nabla \cdot (\varrho \pv) =D_\rho \nabla^2 \varrho,
\label{eq:massconservation}
\\
&\partial_t \pv + \lambda v_0  (\pv \cdot \bm\nabla)  \pv  = \left(\beta |\pv|^2 -\alpha\right)\pv - \frac{v_1}{\varrho_0} \bm \nabla \varrho+ \nu \Delta \pv, \label{eq:a-b}
\end{align}
where $\varrho(\mathbf r,t)$ is the density of active particles that fluctuates around its mean value $\varrho_0$. The polarization field of the material, $\pv(\mathbf r,t)$, denotes the local average orientation of the velocities of the self-propelled units which, when isolated, all move at the same speed $v_0$. The effective viscosity, $\nu$,  the diffusivity, $D_\rho$, and the other (positive) hydrodynamic coefficients $\lambda$, $v_1$, $\alpha$, and $\beta$ have been computed from a number of microscopic models in Refs.~\cite{Bertin2006,Farrell2012,Solon2013,Bricard2013,Frey2015}; $\alpha$ and $\beta$ are the Landau coefficients used to model the spontaneous breaking of rotational symmetry; $v_1$ relates pressure and density. In the SI, we provide a concise introduction to the Toner-Tu model and explain how the left hand side of Eq.~(\ref{eq:a-b}) originates from overdamped dynamics of $\pv$ and not from momentum conservation.

\begin{figure}
\includegraphics[angle=0]{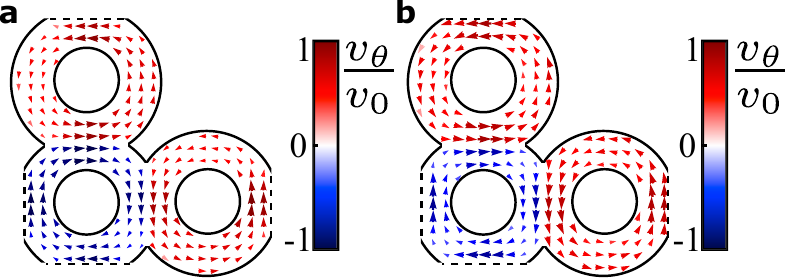}
  \caption{{\bf Steady states of polar active liquids in coupled annular channels.} 
(a) Steady state of a polar active liquid described by the
hydrodynamic Eqs.~(\ref{eq:massconservation},\ref{eq:a-b}), in a confinement geometry based on the Lieb lattice.
Note that the interannular coupling leads to a stable steady-state order reminiscent of either
engaged gears or spins in an antiferromagnet.
The colors indicate the azimuthal component $v_\theta$ of the velocity field (also shown in arrows) around the center of the corresponding annulus.
(b) Steady state of the same liquid simulated using a particle-based model that is described in the SI.
(Dashed lines indicate periodic boundary conditions.)
}
\label{Fig1}
\end{figure}

\begin{figure*}
\includegraphics[angle=0]{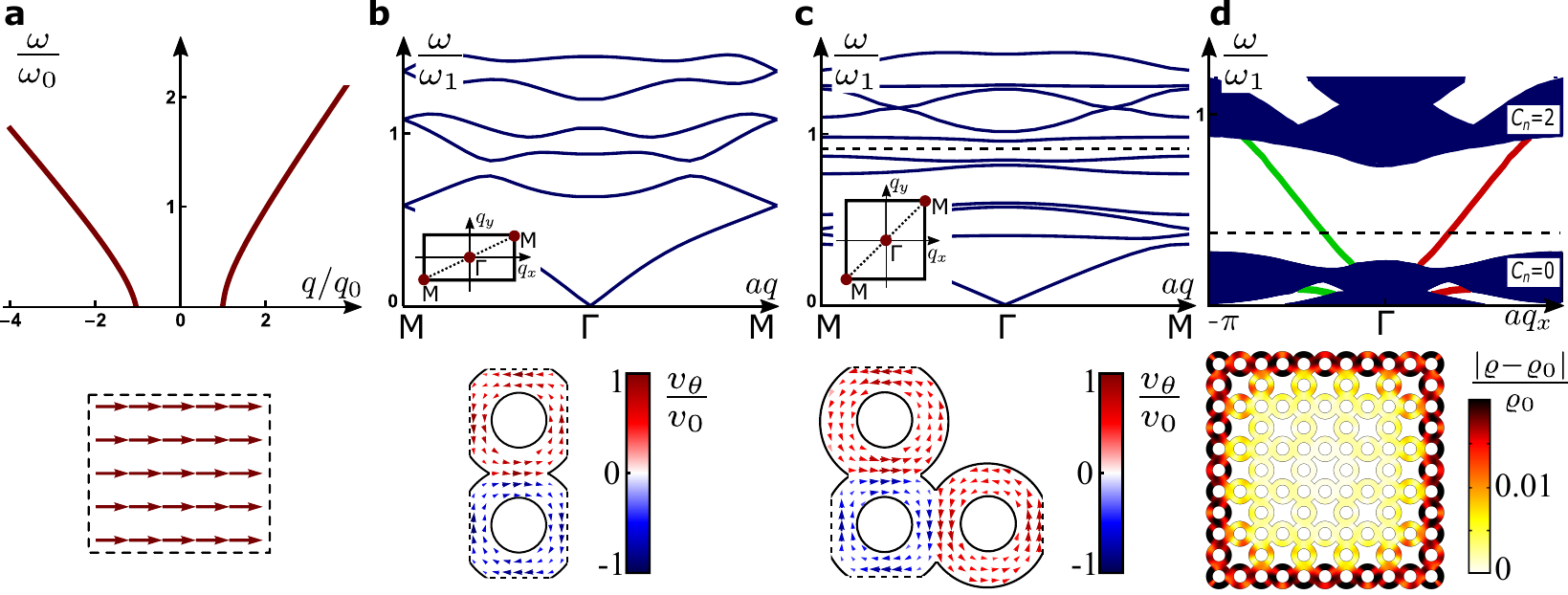}
\caption{
{\bf Dispersion of density waves in active metamaterials either with or without topological edge states.}
(a) Dispersion of longitudinal density waves in an active liquid with a uniform steady-state flow.
For wavenumbers $|q| \gg q_0 \equiv \alpha/c$, these waves have a linear dispersion, reminiscent of pressure waves
in a simple fluid. The spectrum is asymmetric due to the breaking of Galilean invariance by $\pv_0$ (bottom row, a-c: corresponding steady-state flow).
(b) Dispersion of density waves described by Eq.~(\ref{eq:ac-br-ph}) in the square-lattice geometry.
Because the system retains time-reversal symmetry (TRS), bands cross at the point $M$ in the Brillouin zone 
and the bands' Chern numbers are not well defined.
(c) Dispersion of these density waves in the Lieb-lattice geometry.
Due to broken TRS, the bands generically do not cross.
(d) Zoom-in of the dispersion of these waves in a quasi-one-dimensional
waveguide based on the Lieb lattice, with free edges on the top and bottom (also see Fig.~\ref{Fig3}d). The bulk modes (blue) correspond
to the bands in (c). In addition, we observe
chiral topological edge states plotted in red and green colors which indicate state chirality (defined by group velocity $d \omega/dq$) 
and, correspondingly, the edge on which these modes are located.
These states inhabit gaps between bands with well-defined Chern number $C_n \ne 0$.
(below: A density-wave eigenmode of a finite Lieb-lattice sample 
with frequency in this band gap.
The frequency value is indicated by the dashed lines in c-d.)
}
\label{Fig2}
\end{figure*}

Numerically solving Eqs.~\eqref{eq:massconservation} and~\eqref{eq:a-b} in the connected-annuli geometry of Fig.~\ref{Fig1}a, we  find the emergence of a uniform steady chiral flow in each annulus. As this flow is a consequence of spontaneous symmetry breaking, 
left-handed and right-handed orientations are equally likely to occur. These general continuum-mechanics results are confirmed by particle-velocity maps measured from a prototypical microscopic model shown in Fig.~\ref{Fig1}b, see SI. 
As particle velocities align in the region shared between two adjacent annuli, the fluid within these annuli circulates in opposite directions, in analogy with either engaged counter-rotating gears or antiferromagnetic spins. 
Similar behavior was observed in bacterial fluids experiments \cite{Wioland2016} and simulations of agent-based models \cite{Pearce2015}.%

When a homogeneous polar liquid flows through interconnected annuli, the channel geometry determines the mean polarization $\mathbf p_0(\mathbf r)$, which is proportional to the steady-state velocity field. We now elucidate how this emergent spontaneous flow impacts sound propagation. 
We linearize Eqs.~\eqref{eq:massconservation} and \eqref{eq:a-b} deep in the polar liquid phase, in which case both $\alpha$ and $\beta$ are only weakly dependent on $\rho$. We define $\piv (\mathbf r,t)= \mathbf p(\mathbf r,t) - \mathbf p_0(\mathbf r)$ and $\rho(\mathbf r,t) = \varrho(\mathbf r,t) - \varrho_0$, and confirm that density waves propagate over a finite range of wave numbers $q$: $|\alpha| / c \ll q \ll c/ (\nu + D_0) $, where  $c \equiv \sqrt{v_0 v_1}$  sets the magnitude of the speed of sound, see~\cite{Marchetti2013,Toner:2005wy} and SI. In this regime, density fluctuations obey a wave equation that depends on $\mathbf p_0$:
\begin{align}
\label{eq:ac-br-ph}
[\partial_t + \lambda v_0 (\mathbf p_0 \cdot \nabla) ] [\partial_t + v_0 (\mathbf p_0 \cdot \nabla) ] \rho = c^2 \nabla^2 \rho.
\end{align}
Whereas (acoustic) density waves in simple driven fluids~\cite{Khanikaev2015, Yang2015} arise only in systems with inertial dynamics, such waves in polar active liquids survive even in the overdamped limit---in the latter case, these waves originate from Goldstone modes associated with broken rotational symmetry,
see~\cite{Toner:2005wy} and SI.
Fig.~\ref{Fig2}a shows the dispersion relation of density waves for a homogeneous polar liquid uniformly flowing along the $x$-direction ($\mathbf p_0(\mathbf r)= p_0 \hat{x}$). Note that the speed of 
sound depends on the orientation of the wavevector $\mathbf q$ relative to $\mathbf p_0$, because Galilean invariance is broken in Eq.~\eqref{eq:a-b}.

\begin{figure*}
\includegraphics[angle=0]{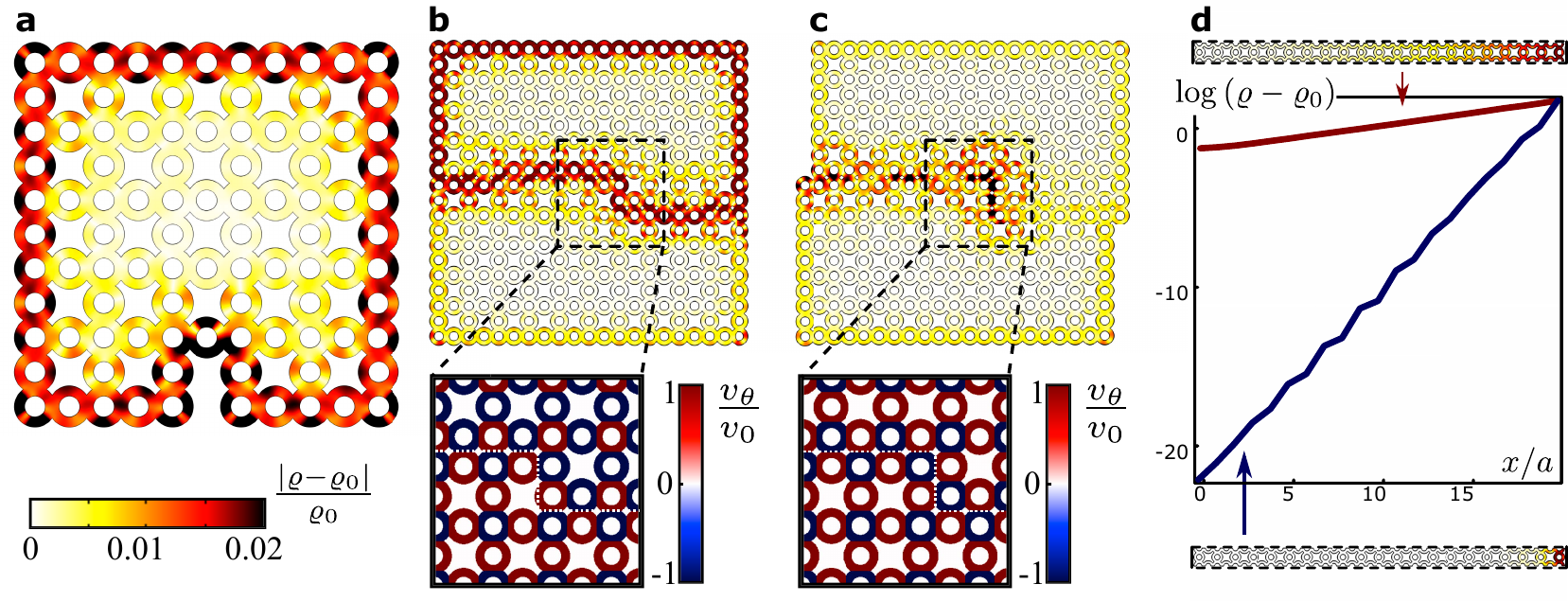}
  \caption{{\bf Topologically protected waveguides composed of an active metamaterial.} 
  (a) The chiral edge state in Fig.~2d is robust against defects along the edge: the state goes around
  a defect instead of backscattering.
  (b) This robustness extends to domain walls separating two different topological phases,
  constructed by taking advantage of antiferromagnetic interannular coupling and deleting a row of the Lieb lattice. 
  Note that the edge state follows the domain wall, no matter the wall's shape. 
  (c) By contrast, along a domain wall that doesn't separate different topological phases,
   the edge states are not robust to backscattering.
   They can scatter off kinks in the wall shape, in the middle of the sample.
   (bottom row, b-c: zoom-ins of the steady-states.)
   (d) Profile of the density within edge states shows that the state decays exponentially into the bulk.
   These exponentially localized edge states are characterized by their penetration depth,
   which can be controlled by changing the flow speed (top: $p_0 = 0.42$, bottom: $p_0 = 0.5$).
}
\label{Fig3}
\end{figure*}

Our design of topological metamaterials exploits (i) microscopic irreversibility induced by activity and (ii) parity-symmetry breaking of the structure. To highlight how the interplay between activity and structural design leads to metamaterials that globally break time reversal symmetry, we contrast two simple geometries of interconnected channels: one based on the square lattice, Fig.~\ref{Fig2}b, and one based on the Lieb lattice, Fig.~\ref{Fig2}c. Solving Eq.~(\ref{eq:ac-br-ph}) numerically in a square lattice geometry (see Methods), we show that the wave spectrum contains degeneracies at the edge of the Brillouin zone where two spectral branches intersect (point M). Note that the corresponding steady-state flow is invariant with respect to simultaneously inverting the arrow of time and performing a lattice translation.
By contrast, the degeneracy at point M is lifted for the Lieb lattice and a gap opens. Unlike the square lattice, the Lieb lattice has an odd number of rings per unit cell and, therefore, a net circulation of steady-state flow. 
Heuristically, the spectral-gap opening stems from the frequency difference between density waves propagating along versus opposite to flow with a non-vanishing net circulation. As a result, a gap opens only for unit cells that are chiral.  
In the limit $v_0 p_0 /c \ll 1$, we rewrite Eq.~\eqref{eq:ac-br-ph} as 
\begin{equation}
\label{eq:pot3}
 \left[(\nabla - i \Av)^2 + \omega^2/c^2 \right] \rho = 0,
\end{equation}
where $\Av \equiv \omega (\lambda +1) v_0 \pv_0/(2 c^2)$,
and note that the steady-state velocity field $v_0 \mathbf p_0$ couples minimally to the wavenumber of the density wave~\cite{Yang2015}.
The emergent chiral flow plays the role of a synthetic gauge field for a charged quantum particle, whereas its curl, the vorticity, acts analogously to a magnetic field that lifts spectral degeneracies.


We establish the topological nature of the band structure corresponding to Eq.~\eqref{eq:ac-br-ph} in the Lieb lattice by calculating (for each band) an integer-valued topological invariant called the Chern number, $C_n$, see Methods and Ref. \cite{Hasan2010} for an introduction.  For almost all of the bands in the spectrum, and for a wide range of values of the mean polarization $p_0$, we find that $C_n \neq 0$, Fig.~\ref{Fig2}c-d.
As $C_n$ is an integer, it cannot vary smoothly from within the sample to the exterior (where $C_n=0$). Therefore, $C_n$ can only change if the band gap closes along the sample edge, locally enabling edge-mode propagation ~\cite{Hasan2010}.
Such edge modes, shown in Fig.~\ref{Fig2}d, are called {\it topologically protected} because they arise from the presence of topological invariants in the bulk, irrespectively of the sample's shape or disorder. As in quantum Hall fluids, the topological edge modes are chiral, i.e., they propagates along a single direction. The chirality of the edge modes reflects the chirality of the flow within the unit cell. The system edge acts as a robust acoustic diode---topological density waves, unlike ordinary waves, propagate unidirectionally along the boundary and do not backscatter even if obstacles or sharp corners are introduced, as demonstrated in Fig.~\ref{Fig3}a.

Similarly, along the boundary between two regions of distinct flow chirality, $C_n$ varies from one integer value to another. 
Therefore, in this region of space, the band gap must vanish, which leads to the existence of topologically protected waves along this interface. 
A topological waveguide can be sculpted in the bulk
by deleting a row of annuli, as in Fig.~\ref{Fig3}b. For this sample, topologically robust density waves propagate through the irregularly shaped domain wall in the bulk of the metamaterial. However, if the domain wall has both a row deletion \emph{and} a half-column displacement, then the chirality of flow does not change across the  wall. Consequently, modes associated with the domain wall are not topologically protected 
and do backscatter in the bulk, as exemplified in Fig.~\ref{Fig3}c.

Whereas the existence of {edge} waves in polar active liquids is topologically protected, their penetration depth {into the bulk} can be tuned by changing the flow speed. As shown in the SI, by considering the minimal coupling form of Eq.~\eqref{eq:pot3} relevant to the motion of density waves in the limit $v_0 p_0 /c \ll 1$, we expect the penetration to be exponential with a penetration depth $\ell$ scaling as $\ell \sim |\Av|^{-1} \sim a c /(v_0 p_0)$, where $a$ is the lattice spacing. We stress that this spatial structure differs from the Gaussian profiles of quantum Hall states that share similar topological properties.
These predictions are in good agreement with the numerical resolution of the full equations of motion: 
as shown in Fig.~\ref{Fig3}d, the penetration of the edge modes is exponential and decreases with the mean-flow speed.

Having explored the phenomenology of chiral states in confined active liquids, we can now compare this realization of a topological metamaterial with those achieved in driven liquids ~\cite{Khanikaev2015, Yang2015}. First, in both cases, to achieve a small penetration depth it is necessary that the speed of flow be appreciable relative to the speed of sound. For a simple fluid, this is a limitation --- driving the fluid at speeds near the speed of sound leads to flow instabilities either in the bulk or in the boundary layer of the fluid. By contrast, for active liquids, the speed of flow $v_0 p_0$ and the speed of sound $c$ are distinct parameters entering the hydrodynamic Eqs.~(\ref{eq:a-b}) and may, in general, be comparable, so that the chiral edge state may be readily observable. 
Second, whereas metamaterials composed of driven fluids require motors at each component to provide the drive, for an active liquid the drive is provided by the particles composing the liquid, whereas the confining channels prescribe the emergent chiral flow. Third, topological density waves in polar active liquids originate from Goldstone modes due to broken rotational symmetry. As a consequence, they can propagate even if particle dynamics are overdamped---paving the way towards colloidal and other soft matter realizations of mechanical topological insulators. 

We examined topological sound in metamaterials based on polar active liquids, but our approach can be applied to wave propagation in other time-reversal-symmetry-broken active systems. Our results epitomize the defining feature of topological active metamaterials: they combine the microscopic irreversibility inherent in active matter with structural design to achieve functionalities absent in passive materials.


\textbf{Acknowledgments}: We thank M. E. Cates, M. C. Marchetti, R. E. Goldstein, J. Paulose, R. Fleury, V. Cheianov, A. Abanov, Z. Yang, and B. Zhang for useful discussions. A.S., B.C.v.Z., and V.V. were funded by FOM, NWO (Vidi grant), and the Delta Institute for Theoretical Physics. D.B. acknowledges support from ANR grant MiTra.

\subsection{Methods}

{\it Chern numbers} --- We establish the topological nature of the active-liquid metamaterial by calculating (for the Lieb-lattice spectrum) an integer-valued topological invariant called the \emph{Chern} number associated with each band, see~\cite{Hasan2010}.
The Chern number $C_n$ is analogous to the Euler characteristic of a closed surface with Gaussian curvature. Using the Gauss-Bonnet theorem, we can compute $C_n$ by integrating a curvature called the \emph{Berry curvature} $B_n(\qv)$ over a closed surface formed by the first Brillouin zone (which by construction is periodic in both directions):
\begin{equation}
C_n \equiv \frac{1}{2 \pi}\int_{\mathrm{BZ}} B_n(\qv)  d\qv,
\end{equation}
 where $ B_n(\qv) \equiv \nabla \times {\cal A}_n(\qv)$, ${\cal A}_n(\qv)\equiv i (\uv^{n}_{\qv})^{\dagger} \cdot (\nabla_{\qv} \uv^{n}_{\qv})$
is the Berry connection calculated from the $\uv^{n}_{\qv}$ eigenstate of band $n$ and wavenumber $\qv$.
For our discrete data set, we use the gauge-choice-independent protocol described in Ref.~\cite{Fukui2013} to efficiently calculate the Chern number using a coarse discretization of the first Brillioun zone.

{\it COMSOL simulations} --- We solve Eq.~(\ref{eq:ac-br-ph}) for both a finite geometry and for a unit cell with Floquet boundary conditions (i.e., periodic boundary conditions with an additional phase factor) using COMSOL Multiphysics finite element analysis simulations on a highly refined mesh.
To obtain the dispersion relations shown in Fig.~2b-c, we perform a sweep through the wavenumbers $(q_x,q_y)$ along the $M\Gamma M$ cut and assign the appropriate phase factors for (Floquet) boundary conditions across the unit cell. Then, we solve the corresponding eigenvalue problem at each wavenumber and plot the corresponding bands. We numerically obtain the solutions in the form of frequencies $\omega_n(\qv)$ as well as the density eigenstates $\tilde{\varrho}(\omega,\qv)$, for which the density waves are $\varrho(x,t) = \tilde{\varrho}(\omega,\qv) e^{i(\omega t - \qv\cdot \xv)}$.
Unless otherwise noted, to obtain good numerical accuracy, we use for the corresponding background flow a simplified model with constant $|\vv| = p_0 v_0 = 0.5 c$, pointed along the azimuthal direction of the corresponding annulus (see visualizations in insets of Fig.3b-c).
In the regions of annular overlap, we patch the flow field using an interpolation that is linear along the $x$-direction, and then normalize the result.
For Fig.~2d, we begin with a quasi-one-dimensional lattice geometry (also see Fig.~3d), and impose a phase factor only along the periodic boundaries in the $x$-direction. Again, the eigenvalues are plotted, and those forming a solid region corresponding to the bulk bands are shaded in blue.
For parts of Figures 2d and 3a-c, we use a finite geometry and plot a single eigenmode located in the band gap that contains topological states.

\pagebreak

\section{Topological sound in active-liquid metamaterials: Supplementary Information}

\vspace{2em}

\setcounter{equation}{0}
\setcounter{figure}{0}

\renewcommand{\theequation}{S\arabic{equation}}
\renewcommand{\figurename}{{\bf Fig. }}
\renewcommand{\thefigure}{{\bf S\arabic{figure}}}
\vspace{2em}

This document is organized as follows: 
We first provide a derivation of the Toner-Tu equations~\cite{Toner1995} introduced in the main text by building on conservation laws and symmetry principles. We also recall how density waves can propagate in polar active fluids despite overdamped microscopic dynamics. 
We then establish  a  rigorous  analogy between
the effect of spontaneous active flow on density waves
and the effect of a gauge field on the wave function of a charged quantum particle. 
Finally, we show how a scaling analysis of the equation for density waves leads
to an estimate for the penetration depth of a topological edge state in an active metamaterial. 

\subsection{Toner-Tu hydrodynamics of polar active liquids}
The hydrodynamic equations of  a {\em passive} polar liquid take into account three slow variables: the usual density, $\rho(\mathbf r,t)$, and velocity, $\mathbf v(\mathbf r,t)$ fields,  as well as a broken-symmetry field, the  polarization $\mathbf p(\mathbf r)$,  defined as the local average of the particle orientations. 
\as{When the polar units that form the liquid propel themselves on a solid surface, momentum is no longer conserved, because the substrate acts as a momentum sink. Such systems are referred to as \emph{dry active matter}, even though the particles may propel in a fluid medium as in, 
e.g., Refs.~\cite{Bausch2010,Bricard2013}. The substrate enables preferential alignment of the particles' velocities with their polar orientation. The  hydrodynamic equations of the resulting polar {\em active}  liquid read:}%
\as{\begin{align}
\partial_t \varrho +\nabla_i(\varrho v_i)& = D^\rho \nabla^2 \rho \label{eq:r}, \\ 
\partial_t (\varrho v_j ) + \nabla_i \left(\varrho v_i v_j \right) & = \nabla_i \sigma_{ij} - \Gamma^v (v_j - v_0 p_j) \label{eq:v}, \\ 
\partial_t p_j +  v_i\nabla_ip_j+\omega_{ji} p_i  & =\nu_1 v_j+ \nu_2 v_{ji} p_i-\Gamma^p  \frac{\delta {\cal H}}{\delta p_j} \label{eq:p},
\end{align}
where we have introduced the symmetric part of the strain-rate tensor $ v_{ji} \equiv \frac{1}{2}\left( \partial_j v_i + \partial_i v_j \right)$ and the vorticity tensor $ \omega_{ji} \equiv \frac{1}{2}\left( \partial_j v_i - \partial_i v_j \right)$. 
Note that the components of the velocity vector $v_i$ for this one-fluid model are the coarse-grained velocities of the self-propelled particles composing the active liquid and \emph{not} of the potential surrounding fluid (e.g., air or solvent). }
The first (continuity) equation reflects mass conservation and includes a diffusive term $D^\rho \nabla^2 \rho$. The
second equation includes the liquid stress tensor $\sigma$ as well as an active frictional term proportional to $\Gamma^v$. 
This terms differentiates Eq.~\eqref{eq:v} from the usual Navier-Stokes equations as it explicitly  breaks momentum conservation. For the sake of simplicity, we consider here a linear coupling between the velocity and the polarization (the hydrodynamic coefficient $v_0$ has the dimensions of a speed and scales with the speed of an isolated active particle).
\as{Equation~(\ref{eq:p}) describes the relaxational 
dynamics of the polarization, see~\cite{Marchetti2013}.  The left-hand side of Eq.~(\ref{eq:p}) contains the comoving (2nd term), corotational (3rd term) time-derivative of the polarization. The right-hand side of Eq.~(\ref{eq:p}) includes the effective Hamiltonian ${\cal H}$ and the dissipative coefficient $\Gamma^p$ along with two frictional terms. The first frictional term in Eq.~(\ref{eq:p}) contains the friction coefficient $\nu_1$ and describes the friction between particle and substrate---this terms is responsible for the ``weathercock effect,'' i.e., the polar particles' local alignment with the flow see, e.g., Refs.~\cite{SoodRamaswamy2014, Brotto}. The second friction term in Eq.~(\ref{eq:p}) contains the friction coefficient $\nu_2$ and originates from the friction between an individual polar particle and the surrounding active fluid (itself composed of polar particles). The sign and the magnitude of $\nu_2$ controls the strength of alignment of the particle polarization with the local elongation (or compression) axis of the flow. }

We can also consider Eqs.~\eqref{eq:r}, \eqref{eq:v}, and \eqref{eq:p}  in the limit for which the frictional $\Gamma^v$ term
dominates Eq.~(\ref{eq:v}). In this overdamped limit, Eq.~(\ref{eq:v}) reduces to  a constraint equation, $\vv = v_0 \pv$, and the hydrodynamics is fully captured by mass conservation and the dissipative dynamics of the polarization field. A gradient expansion of ${\cal H}$ then yields~\cite{Toner1995,Toner:2005wy}:
\begin{align}
&\partial_t \varrho +  v_0\bm\nabla \cdot (\varrho \pv) =D_0 \nabla^2 \varrho,
\label{eq:massconservation1}
\\
&\partial_t \pv + \lambda v_0  \pv \cdot \bm \nabla  \pv  = \left(\beta |\pv|^2 -\alpha\right)\pv - \frac{v_1}{\varrho_0} \bm \nabla \varrho+ \nu \Delta \pv + \lambda_2 v_0 \nabla |\pv|^2 - \lambda_2 v_0 \pv (\nabla \cdot \pv), \label{eq:a-b1}
\end{align}
where all of the hydrodynamic coefficients depends \emph{a priori} on the local density.
Note that whereas for a system with Galilean invariance, $\lambda = 1$ and $\lambda_2 = 0$, for the polar active liquid, which lacks this symmetry, these parameters
may be arbitrary. Studies of realistic microscopic models have found $\lambda$ to be positive, less than, and of order $1$, and for the numerical computations performed in this work, we assume $\lambda = 0.8$~\cite{Bertin2006,Farrell2012,Bricard2013}.
The lack of Galilean invariance as well as momentum conservation leads to the $\alpha$ and $\beta$ terms in Eq.~(\ref{eq:a-b1}), 
which suggests a preference for either zero or nonzero velocity---depending on the sign of $\alpha$.  In the article, for the sake of simplicity we focus on the case in which the $\lambda_2$ terms are negligible.
In this case, Eqs.~(\ref{eq:massconservation1},\ref{eq:a-b1}) reduce to Eqs.~(1,2) in the main text, and are reproduced here:
\begin{align}
&\partial_t \varrho +  v_0\bm\nabla \cdot (\varrho \pv) =D_\rho \nabla^2 \varrho,
\label{eq:massconservation2}
\\
&\partial_t \pv + \lambda v_0  (\pv \cdot \bm\nabla)  \pv  = \left(\beta |\pv|^2 -\alpha\right)\pv - \frac{v_1}{\varrho_0} \bm \nabla \varrho+ \nu \Delta \pv, \label{eq:a-b2}
\end{align}
Eqs.~(\ref{eq:massconservation2},\ref{eq:a-b2}) are sufficient to capture the phenomena associated
with linear density waves in a polar active liquid relevant to our analysis.
In that limit, the polarization field itself defines the fluid velocity, so that 
the coupling between the polarization field and the density gradient has an effect analogous
to that of a pressure gradient in an equilibrium liquid.

\subsection{Linear density waves for Toner-Tu liquids}
For the case $\alpha > 0$, the Toner-Tu equations result in a steady state of the fluid with spontaneous flow in the bulk, such that
$p_0^2 \equiv |\pv_0|^2 = \alpha/\beta$. Although in the bulk, the spontaneous flow direction $\hat{\pv}_0$ could be arbitrary,
in physical realizations of active liquids, the boundaries fix $\hat{\pv}_0$. For example, in an open channel, $\hat{\pv}_0$ is parallel to the channel walls.
In the Lieb lattice geometry, we have solved Eqs.~(\ref{eq:massconservation2}) and (\ref{eq:a-b2}) for a sufficient time for the dynamics to relax to a steady state.
We find that this steady state, plotted in Fig.~1a (also see~\ref{FigS2}), has the features of the spatial profile observed from our particle-based simulations,
although the particle-based simulations lead to a smoother profile, Fig.~1b. 

In the analysis performed for the density wave computations, we take a particularly simple form of the steady state,
based on the profile we observe. We postulate the polarization has magnitude unity everywhere and is oriented
azimuthally, i.e., perpendicular to the vector connecting the position of the fluid to the nearest annulus center. 
In the regions of overlap between annuli, we linearly interpolate between the two annular flow profiles. This
spatial profile is plotted in a large sample in Fig.~3a of the main text.

Thus, given a spontaneous steady-state flow field $\pv_0$, we expand $\piv (\mathbf r,t)= \mathbf p(\mathbf r,t) - \mathbf p_0(\mathbf r)$ and $\rho(\mathbf r,t) = \varrho(\mathbf r,t) - \varrho_0$ to find
\begin{align}
\label{eqS:ac-br-ph}
&\partial_t \rho + (v_0 \pv_0 \cdot \nabla) \rho = - v_0 \rho_0 \nabla \cdot \vv + D_0 \nabla^2 \rho \\
&\partial_t \vv + \lambda v_0 (\pv_0 \cdot \nabla) \vv = - (v_1/\rho_0) \nabla \rho - 2 \alpha (\vv \cdot \hat{\pv}_0)\hat{\pv}_0 + \nu \nabla^2 \vv. \nonumber
\end{align}
For the case of propagating waves, the right-hand side can be decomposed as the sum of a dominant anti-Hermitian matrix
that governs wave dispersion and a perturbatively small Hermitian matrix that governs wave attenuation.
As we are interested in the behavior of an active fluid deep in the ordered phase, 
we have assumed $\alpha$ to be constant. The $\rho$ dependance of $\alpha$, which leads to additional dissipative terms,
would be most significant near the phase transition from an isotropic to a flowing steady state.
There are two notable differences for the propagation of density waves in an active liquid compared to a simple fluid: 
(1) the $\alpha$ term acts as an additional dissipative term for sound in an active liquid, and (2) one of the convection terms contains the coefficient $\lambda$~($\ne 1$).  Due to this second difference, the equation of motion can no longer be ``Galilean boosted'' into a different
reference frame by replacing the lab-frame derivative $\partial_t$ by a convective derivative.

To closely examine the mode structure in Eqs.~(\ref{eqS:ac-br-ph}), 
we split the vector $\vv$ into components $v_\parallel$ and $v_\perp$, respectively parallel and perpendicular 
to $\hat{\pv}_0$. Note that due to the confinement of the active liquid inside a channel, 
we can assume that the density waves only propagate along the channel
and, therefore, the derivatives of $\pv$ and $\rho$ along the direction perpendicular to $\hat{\pv}_0$ can be ignored.
Under this assumption, Eqs.~(\ref{eqS:ac-br-ph}) reduce to:
\begin{align}
&\partial_t \rho + v_0 p_0 \partial_\parallel \rho = - v_0 \rho_0 \partial_\parallel v_\parallel + D_0 \partial_\parallel^2 \rho, \label{eq:ac-r}\\
&\partial_t v_\parallel + \lambda p_0 \partial_\parallel v_\parallel = - (v_1/\rho_0) \partial_\parallel \rho - 2 \alpha v_\parallel + \nu \partial_\parallel^2 v_\parallel, \label{eq:c} \\
&\partial_t v_\perp + \lambda p_0 \partial_\parallel v_\perp = \nu \partial_\parallel^2 v_\perp. \label{eq:ac-perp}
\end{align}
Let us now consider  Eqs.~(\ref{eq:ac-r}-\ref{eq:c}) for the density and the longitudinal velocity modes in an active liquid.
Ignoring, for now, the effects of the flow, we can calculate the dispersion relation for density waves in active liquids in the limit $q \ll c / (D_0 + \nu)$, 
where $q$ is the wavenumber of the density wave and $c \equiv \sqrt{v_0 v_1}$ is the speed of sound. The frequency is then given by $\omega(q) = i |\alpha| + \sqrt{q^2 v_0 v_1 - \alpha^2} + i (D_0 + \nu) q^2 / 2$ (also see Fig.~2a).
Whereas the real component of $\omega$ governs the propagation of sound waves,
the imaginary component governs their dissipation.
Due to spontaneous flow, the sound wave frequency may, generically,
have a real component, which is plotted in Fig.~1a of the main text.
Furthermore, in the limits $q \gg |\alpha|/c$, the imaginary component will always be much smaller than the real component. 
Thus, in the regime $|\alpha|/c \ll q \ll c/(D_0 + \nu)$,
there exist longitudinal density waves that propagate and decay slowly in the ordered active liquid.
Provided we are in this regime, for phenomena on sufficiently short time scales, we may ignore
the dissipative terms and concentrate on the wave-like solutions to the equations of motion.

\subsection{Analogy with Schr\"{o}dinger equation}
Note that when the dissipative components of the density wave equation can be neglected, Eq.~(\ref{eqS:ac-br-ph})
may be recast as a single wave equation. By applying the convective derivative $\partial_t + \lambda v_0 (\mathbf p_0 \cdot \nabla)$
to the continuity equation, Eq.~\eqref{eq:ac-r}, and then substituting the velocity equation of motion, Eq.~\eqref{eq:c}, one obtains:
\begin{align}
\label{eq:wave}
[\partial_t + \lambda v_0 (\mathbf p_0 \cdot \nabla) ] [\partial_t + v_0 (\mathbf p_0 \cdot \nabla) ] \rho = c^2 \nabla^2 \rho.
\end{align}
The eigenvalue problem for the above wave equation has solutions in terms of the frequency $\omega$ of a time-dependent
oscillation $\rho(\xv,t) = \tilde{\rho}(\xv,\omega) e^{i \omega t}$. 
The corresponding equation has the form, provided that $M \equiv v_0 p_0 / c \ll 1$,
\begin{equation}
\label{eq:pot2}
 \left[c^2 \nabla^2 + \omega^2 - i \omega (\lambda + v_0) \pv_0\cdot \nabla \right] \tilde{\rho} = 0,
\end{equation}
or, 
\begin{equation}
\label{eqS:pot3}
 \left[(\nabla - i \Av)^2 + \omega^2/c^2 \right] \tilde{\rho} = 0,
\end{equation}
where $\Av \equiv \omega (\lambda +1) v_0 \pv_0/(2 c^2)$.
This shows that the velocity field $v_0 \pv_0$ acts as an effective vector potential for the propagation of density waves.

\subsection{Scaling argument for penetration depth}
From the form of Eq.~(\ref{eqS:pot3}), we can deduce the following scaling argument for the penetration depth
of a topological edge state in the relevant limit $v_0 p_0 /c \ll 1$. Consider the first term, $(\nabla - i \Av)^2$,
which shows the minimal coupling between density gradients and spontaneous flow~\cite{Yang2015}.
The penetration depth is a lengthscale that originates from density gradients and therefore scales as
$\ell \sim \Av^{-1}$. Furthermore, $\Av \sim v_0 p_0 \omega/ c^2$ and depends on a characteristic frequency $\omega \sim c/a$,
where $c$ is the speed of sound and $a$ is a characteristic lengthscale of the material, i.e., the lattice spacing.
In addition, $\Av$ is approximately the same from one unit cell to the next.
Combining these scaling relations, $\ell \sim a c /(v_0 p_0)$. The length $\ell$ diverges
as the flow velocity goes to zero and, therefore, as the material loses its bulk bandgap.

We also note that we expect and observe topological
edge states to be localized near the edge with an \emph{exponential} profile, see Fig.~3d in the main text.
To see why we expect Eq.~(\ref{eqS:pot3}) to lead to exponentially localized states,
note that if we assume $\rho \sim f(\xv/\ell)$, 
and the scaling law derived above for $\ell$, $\ell \sim a c/ (v_0 p_0)$,
Eq.~(\ref{eqS:pot3}) predicts $f^{\prime\prime} \sim f$, with a dimensionless proportionality constant.
An exponential profile satisfies this approximate scaling form.
Such a profile is a consequence of the fact that $\Av$ does not vary over lengthscales larger than a unit cell, 
an argument that relies on the metamaterial structure of the topological state.
By contrast, in the quantum Hall fluid, the frequency scale depends on the field strength and $\Av$ varies over large distances,
which leads to both a Gaussian profile of states in a Landau level
as well as a different scaling for the penetration depth~\cite{Landau3}.

\subsection{Particle-based model}
\begin{figure*}
\includegraphics[angle=0]{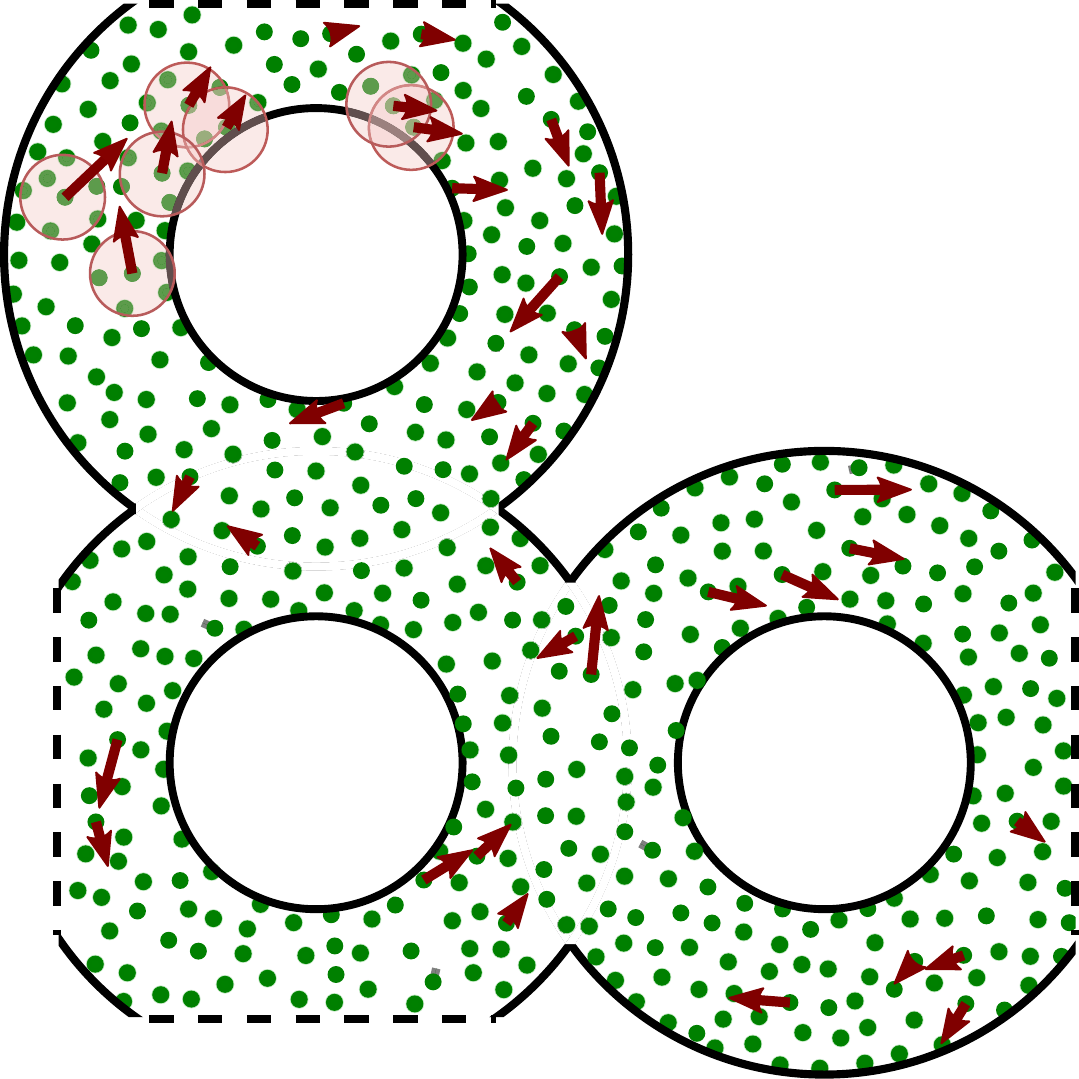}
  \caption{ One configuration of the particle-based simulation in a periodic geometry based on the Lieb lattice. 
  For each particle, the radius of the short-range repulsive interaction is indicated in green.
  For a few chosen particles, the radius of the longer-range alignment interaction is indicated in pink.
  For some particle, their instantaneous velocity is indicated by a red arrow.}
\label{FigS1}
\end{figure*}

\as{We used a particle-based model as an illustrative example to check the steady-state flow that we obtained from the Toner-Tu equations.
We emphasize that the conclusions obtained in the main text are based on the continuum Toner-Tu equations, which form a description that has a more general applicability than the specific particle-based model presented below.
We choose a continuous-time model that includes Vicsek-like alignment interactions and repulsive interactions that prevent clustering~\cite{Vicsek1995, Levine2000}.}
The position $\mathbf x_i$ and velocity $\mathbf v_i$ of the $i^{\rm th}$ particle is evolved using a symplectic Euler integrator for Newton's laws of motion with the force term
\begin{align}
    \begin{split}
        \Fv_i = m \dot{ \vv}_i = &- \gamma \vv_i + F_0 \left( \hat{\vv}_i + \sum_{{\left< {\bf x}_i,{\bf x}_j \right>}} \hat{\vv}_j /N \right) \\
        &+ \sum_{k} \nabla_i U(|\xv_i - \xv_k|) + \sqrt{2 \gamma k_B T} \hat{\Zv}_i(t),
    \end{split}
\end{align}
where $\gamma$ is a friction coefficient, $m$ is the particle mass, and $F_0$ is the active force such that $v_0 = F_0 / \gamma$. 
The neighbors in the $F_0$ term are denoted as $\left< {\bf x}_i,{\bf x}_j \right>$ and include all particles ${\bf x}_j$ within a distance $R$ ($= a/20$) of ${\bf x}_i$, \as{see Fig.~\ref{FigS1}}.
We use a Yukawa potential $U(r)$:
\begin{align}
    U(r) = \frac{b}{r e^{\kappa r}}
\end{align}
to account for excluded-volume effects,
where $\kappa^{-1} = R/6$ sets the repulsion range, and $b=4 \times 10^3 F_0/\kappa^2$ sets the Yukawa coupling constant.
The white-noise stochastic forcing term $\hat{\Zv}_i(t)$, where $\langle \hat{\Zv}_i(t) \hat{\Zv}_i(t^\prime)\rangle = \delta(t - t^\prime)$, mimics thermal fluctuations.
The temperature is set by $k_B T = 10^{-3} b \kappa = 2 \times 10^{-2} m v_0^2$. \as{The nonlinear forcing term $F_0 \hat{\vv}$, where $\hat{\vv} \equiv \vv/|\vv|$, breaks the equilibrium fluctuation-dissipation relation for this far-from-equilibrium system.} \as{The overdamped limit is defined as the regime for which the velocity relaxation time $\tau_{\rm p}\equiv m/\gamma$ is much smaller that the characteristic oscillation time $\tau_o$ associated with the interaction potential: $\tau_o\equiv \sqrt{m b^{-1}\rho^{-3}}$, where $\rho$ is the particle density. }
Time integration is done using the following time step $\Delta t = 10^{-5} m/\gamma$, where $m$ is the mass of an individual particle.

Both the square and Lieb lattices have lattice spacing $a=120 \kappa^{-1}$ and are implemented by confining particles in overlapping annuli.
A single annulus has an inner radius $R_{\text{in}} = \tfrac{3}{10} a$ and $R_{\text{out}}= 2 R_{\text{in}}$.
The confining boundaries are implemented using a steep one-sided harmonic repulsive potential $\tfrac{1}{2} k_w x^2$ with $k_w = 3.14 \times 10^5 \gamma^2/m$ experienced by all particles.
The area fraction of particles is
\begin{align}
    \phi = \frac{N R^{2}}{R_{\text{out}}^2-R_{\text{in}}^2} \approx 6.17,
\end{align}
where $N$ is the number of particles per annulus. (We choose units in which $m =1$, $a =6$, and $a / v_0 = 60$.)

\begin{figure*}
\includegraphics[angle=0]{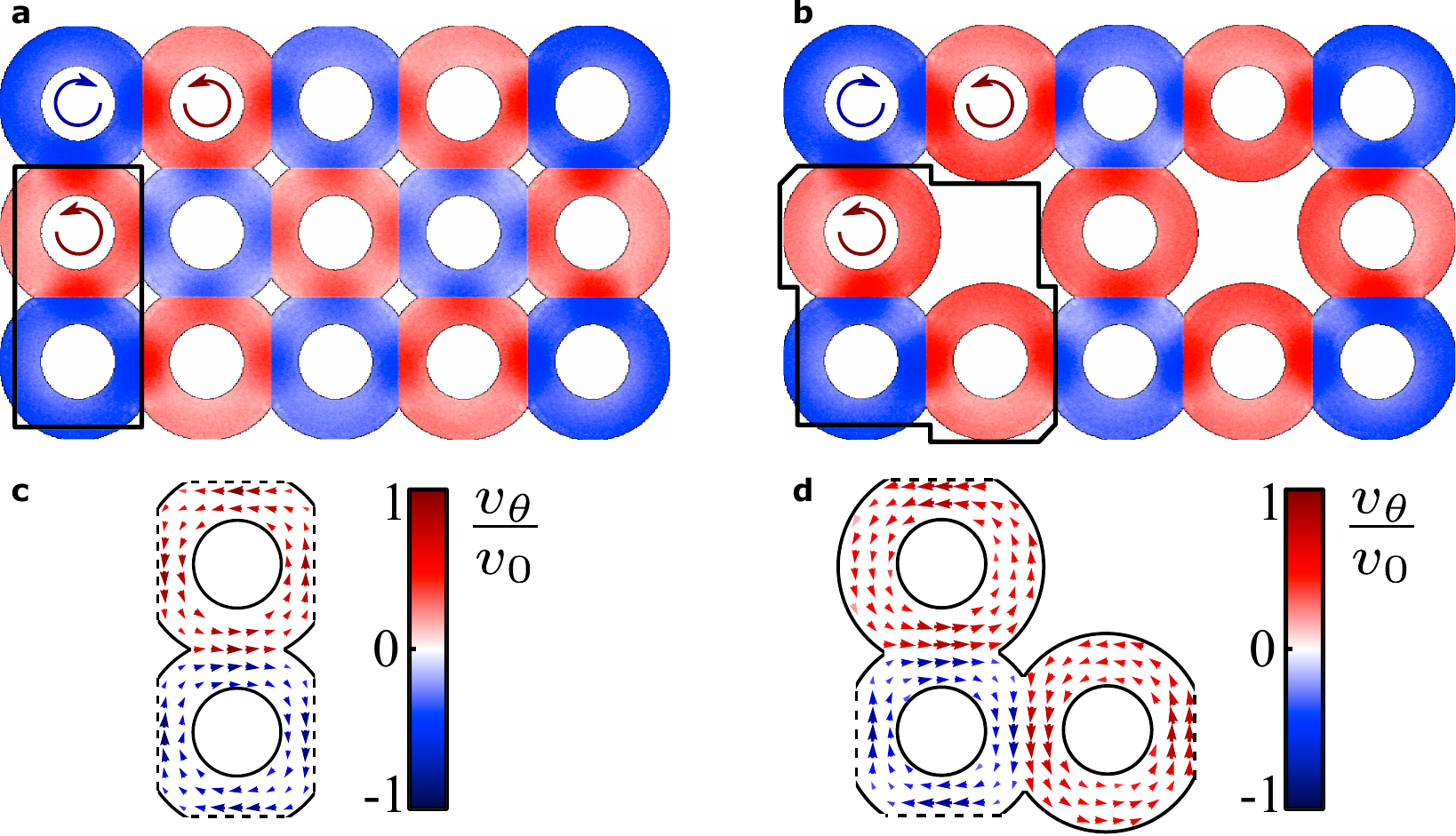}
  \caption{
The steady-state for the flow of a polar active liquid in active metamaterials for (a,c) square lattice and (b,d) topological Lieb lattice, all obtained from molecular dynamics (MD) simulations. 
For both geometries we also reproduce parts of Fig.~2 of the main text, 
showing the spatially-dependent flow within a single unit cell (c,d), with the color
 scale for the azimuthal component of the flow field.}
\label{FigS2}
\end{figure*}

\begin{figure*}
\includegraphics[angle=0]{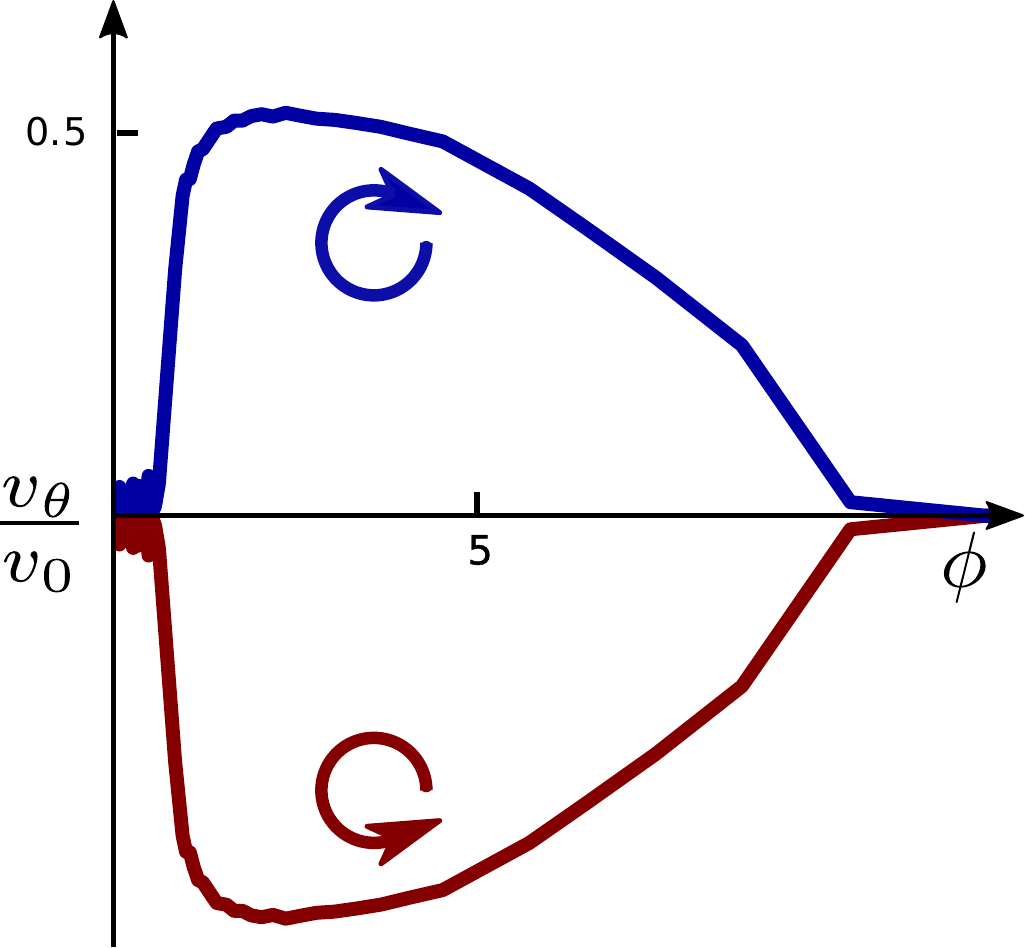}
  \caption{(a) The average normalized azimuthal component $v_\theta/v_0$ (measured relative to the center of each annulus) 
as a function of particle density $\phi$ in the Lieb lattice geometry measured relative to the (large) radius of the alignment interaction (see Methods, Fig.~\ref{FigS1}).}
\label{FigS3}
\end{figure*}

In the steady state, the flow for the $k$-th annulus is measured by the azimuthal component of the velocity field
\begin{align}
v_\theta= \frac{1}{v_0} \left\langle \sum_{i=1}^{N} \vv_i \cdot \hat{\theta}_k \right\rangle,
\end{align}
where $\hat{\theta}_k$ is the azimuthal unit vector around the annulus center,
 and $\langle \dots \rangle$ denotes a time average over $8000$ configuration, with $1000$ timesteps between subsequent configurations.
$v_\theta = \pm 1$ for an ideally flowing system; the sign indicates flow chirality. \as{Two examples of steady states are plotted in Fig.~\ref{FigS2}
and the dependence of $v_\theta$ on $\phi$ within the Lieb lattice is plotted in Fig.~\ref{FigS3}. At low area fraction,
the particles undergo the alignment transition for their velocities, whereas at high area fraction they jam. 
The flowing steady-state occurs over a wide range of intermediate area fractions.}

\clearpage


\begin{thebibliography}{10}
\expandafter\ifx\csname url\endcsname\relax
  \def\url#1{\texttt{#1}}\fi
\expandafter\ifx\csname urlprefix\endcsname\relax\def\urlprefix{URL }\fi
\providecommand{\bibinfo}[2]{#2}
\providecommand{\eprint}[2][]{\url{#2}}

\bibitem{Bausch2010}
\bibinfo{author}{Schaller, V.}, \bibinfo{author}{Weber, C.},
  \bibinfo{author}{Semmrich, C.}, \bibinfo{author}{Frey, E.} \&
  \bibinfo{author}{Bausch, A.~R.}
\newblock \bibinfo{title}{{Polar patterns of driven filaments}}.
\newblock \emph{\bibinfo{journal}{Nature}} \textbf{\bibinfo{volume}{467}},
  \bibinfo{pages}{73--77} (\bibinfo{year}{2010}).

\bibitem{Bricard2013}
\bibinfo{author}{Bricard, A.}, \bibinfo{author}{Caussin, J.-B.},
  \bibinfo{author}{Desreumaux, N.}, \bibinfo{author}{Dauchot, O.} \&
  \bibinfo{author}{Bartolo, D.}
\newblock \bibinfo{title}{{Emergence of macroscopic directed motion in
  populations of motile colloids.}}
\newblock \emph{\bibinfo{journal}{Nature}} \textbf{\bibinfo{volume}{503}},
  \bibinfo{pages}{95--98} (\bibinfo{year}{2013}).

\bibitem{Granick2016}
\bibinfo{author}{Yan, J.} \emph{et~al.}
\newblock \bibinfo{title}{{Reconfiguring active particles by electrostatic
  imbalance}}.
\newblock \emph{\bibinfo{journal}{Nat.~Mater.}} \textbf{15}, 1095--1099 (2016).

\bibitem{Deseigne2010}
\bibinfo{author}{Deseigne, J.}, \bibinfo{author}{Dauchot, O.} \&
  \bibinfo{author}{Chat{\'e}, H.}
\newblock \bibinfo{title}{{Collective Motion of Vibrated Polar Disks}}.
\newblock \emph{\bibinfo{journal}{Phys.~Rev.~Lett.}} \textbf{\bibinfo{volume}{105}}, \bibinfo{pages}{098001}
  (\bibinfo{year}{2010}).

\bibitem{SoodRamaswamy2014}
\bibinfo{author}{Kumar, N.}, \bibinfo{author}{Soni, H.},
  \bibinfo{author}{Ramaswamy, S.} \& \bibinfo{author}{Sood, A.~K.}
\newblock \bibinfo{title}{{Flocking at a distance in active granular matter}}
\newblock \emph{\bibinfo{journal}{Nat.~Commun.}}~\textbf{\bibinfo{volume}{5}}, 
\bibinfo{pages}{4688} (\bibinfo{year}{2014}).

\bibitem{Marchetti2013}
\bibinfo{author}{Marchetti, M.~C.} \emph{et~al.}
\newblock \bibinfo{title}{{Hydrodynamics of soft active matter}}.
\newblock \emph{\bibinfo{journal}{Rev.~Mod.~Phys.}}
  \textbf{\bibinfo{volume}{85}}, \bibinfo{pages}{1143--1189}
  (\bibinfo{year}{2013}).
  
\bibitem{Vicsek_review}
\bibinfo{author}{Vicsek, T.} \& \bibinfo{author}{Zafeiris, A.}
\newblock \bibinfo{title}{{Collective motion}}.
\newblock \emph{\bibinfo{journal}{Phys.~Rep.}}
  \textbf{\bibinfo{volume}{517}}, \bibinfo{pages}{71--140}
  (\bibinfo{year}{2012}).

\bibitem{Toner:2005wy}
\bibinfo{author}{Toner, J.}, \bibinfo{author}{Tu, Y.} \&
  \bibinfo{author}{Ramaswamy, S.}
\newblock \bibinfo{title}{{Hydrodynamics and phases of flocks}}.
\newblock \emph{\bibinfo{journal}{Ann.~Phys.}}
  \textbf{\bibinfo{volume}{318}}, \bibinfo{pages}{170--244}
  (\bibinfo{year}{2005}).

\bibitem{Wioland2016}
\bibinfo{author}{Wioland, H.}, \bibinfo{author}{Woodhouse, F.~G.},
  \bibinfo{author}{Dunkel, J.} \& \bibinfo{author}{Goldstein, R.~E.}
\newblock \bibinfo{title}{{Ferromagnetic and antiferromagnetic order in
  bacterial vortex lattices}}.
\newblock \emph{\bibinfo{journal}{Nat.~Phys.}}  
\textbf{12}, 341--345
(\bibinfo{year}{2016}).

\bibitem{Bricard2015}
\bibinfo{author}{Bricard, A.} \emph{et~al.}
\newblock \bibinfo{title}{{Emergent vortices in populations of colloidal
  rollers}}.
\newblock \emph{\bibinfo{journal}{Nat.~Commun.}}
  \textbf{\bibinfo{volume}{6}}, \bibinfo{pages}{7470} (\bibinfo{year}{2015}).

\bibitem{Yang2015}
\bibinfo{author}{Yang, Z.} \emph{et~al.}
\newblock \bibinfo{title}{{Topological Acoustics}}.
\newblock \emph{\bibinfo{journal}{Phys.~Rev.~Lett.}}
  \textbf{\bibinfo{volume}{114}}, \bibinfo{pages}{114301}
  (\bibinfo{year}{2015}).

\bibitem{Khanikaev2015}
\bibinfo{author}{Khanikaev, A.~B.}, \bibinfo{author}{Fleury, R.},
  \bibinfo{author}{Mousavi, S.~H.} \& \bibinfo{author}{Alu, A.}
\newblock \bibinfo{title}{{Topologically robust sound propagation in an
  angular-momentum-biased graphene-like resonator lattice}}.
\newblock \emph{\bibinfo{journal}{Nat.~Commun.}} \textbf{\bibinfo{volume}{6}}, 8260
  (\bibinfo{year}{2015}).

\bibitem{Hasan2010}
\bibinfo{author}{Hasan, M.~Z.} \& \bibinfo{author}{Kane, C.~L.}
\newblock \bibinfo{title}{{Colloquium: Topological insulators}}.
\newblock \emph{\bibinfo{journal}{Rev.~Mod.~Phys.}}
  \textbf{\bibinfo{volume}{82}}, \bibinfo{pages}{3045--3067}
  (\bibinfo{year}{2010}).

\bibitem{Kane2014}
\bibinfo{author}{Kane, C.~L.} \& \bibinfo{author}{Lubensky, T.~C.}
\newblock \bibinfo{title}{{Topological boundary modes in isostatic lattices}}.
\newblock \emph{\bibinfo{journal}{Nat.~Phys.}}
  \textbf{\bibinfo{volume}{10}}, \bibinfo{pages}{39--45}
  (\bibinfo{year}{2013}).

\bibitem{Chen2014}
\bibinfo{author}{Chen, B.~G.}, \bibinfo{author}{Upadhyaya, N.} \&
  \bibinfo{author}{Vitelli, V.}
\newblock \bibinfo{title}{{Nonlinear conduction via solitons in a topological
  mechanical insulator}}.
\newblock \emph{\bibinfo{journal}{Proc.~Natl.~Acad.~Sci.~USA}} 
\textbf{\bibinfo{volume}{111}}, \bibinfo{pages}{13004--13009}
  (\bibinfo{year}{2014}).

\bibitem{Paulose2015}
\bibinfo{author}{Paulose, J.}, \bibinfo{author}{Chen, B.~G.} \&
  \bibinfo{author}{Vitelli, V.}
\newblock \bibinfo{title}{{Topological modes bound to dislocations in
  mechanical metamaterials}}.
\newblock \emph{\bibinfo{journal}{Nat.~Phys.}} \textbf{\bibinfo{volume}{11}}
\bibinfo{pages}{153--156}
 (\bibinfo{year}{2015}).

\bibitem{Prodan2009}
\bibinfo{author}{Prodan, E.} \& \bibinfo{author}{Prodan, C.}
\newblock \bibinfo{title}{{Topological phonon modes and their role in dynamic
  instability of microtubules}}.
\newblock \emph{\bibinfo{journal}{Phys.~Rev.~Lett.}}
  \textbf{\bibinfo{volume}{103}}, \bibinfo{pages}{248101}
  (\bibinfo{year}{2009}).

\bibitem{Nash2015}
\bibinfo{author}{Nash, L.~M.} \emph{et~al.}
\newblock \bibinfo{title}{{Topological mechanics of gyroscopic metamaterials.}}
\newblock \emph{\bibinfo{journal}{Proc.~Natl.~Acad.~Sci.~USA}} \textbf{\bibinfo{volume}{112}},
  \bibinfo{pages}{14495--14500} (\bibinfo{year}{2015}).

\bibitem{Susstrunk2015}
\bibinfo{author}{Susstrunk, R.} \& \bibinfo{author}{Huber, S.~D.}
\newblock \bibinfo{title}{{Observation of phononic helical edge states in a
  mechanical topological insulator}}.
\newblock \emph{\bibinfo{journal}{Science}} \textbf{\bibinfo{volume}{349}},
  \bibinfo{pages}{47--50} (\bibinfo{year}{2015}).

\bibitem{Kariyado2015}
\bibinfo{author}{Kariyado, T.} \& \bibinfo{author}{Hatsugai, Y.}
\newblock \bibinfo{title}{{Manipulation of Dirac Cones in Mechanical
  Graphene}}.
\newblock \emph{\bibinfo{journal}{Sci.~Rep.}} \textbf{\bibinfo{volume}{5}},
  \bibinfo{pages}{18107} (\bibinfo{year}{2015}).

\bibitem{Wang2015}
\bibinfo{author}{Wang, P.}, \bibinfo{author}{Lu, L.} \&
  \bibinfo{author}{Bertoldi, K.}
\newblock \bibinfo{title}{{Topological Phononic Crystals with One-Way Elastic
  Edge Waves.}}
\newblock \emph{\bibinfo{journal}{Phys.~Rev.~Lett.}}
  \textbf{\bibinfo{volume}{115}}, \bibinfo{pages}{104302}
  (\bibinfo{year}{2015}).

\bibitem{He2016}
\bibinfo{author}{He, C.} \emph{et~al.}
\newblock \bibinfo{title}{{Acoustic topological insulator and robust one-way
  sound transport}}.
\newblock \emph{\bibinfo{journal}{Nat.~Phys.}}  (\bibinfo{year}{2016}).

\bibitem{Toner1995}
\bibinfo{author}{Toner, J.} \& \bibinfo{author}{Tu, Y.}
\newblock \bibinfo{title}{{Long-Range Order in a Two-Dimensional Dynamical
  XY-Model: How Birds Fly Together}}.
\newblock \emph{\bibinfo{journal}{Phys.~Rev.~Lett.}}
  \textbf{\bibinfo{volume}{75}}, \bibinfo{pages}{4326--4329}
  (\bibinfo{year}{1995}).

\bibitem{Bertin2006}
\bibinfo{author}{Bertin, E.}, \bibinfo{author}{Droz, M.} \&
  \bibinfo{author}{Gregoire, G.}
\newblock \bibinfo{title}{{Boltzmann and hydrodynamic description for
  self-propelled particles}}.
\newblock \emph{\bibinfo{journal}{Phys.~Rev.~E}}
  \textbf{\bibinfo{volume}{74}}, \bibinfo{pages}{022101}
  (\bibinfo{year}{2006}).

\bibitem{Farrell2012}
\bibinfo{author}{Farrell, F.}, \bibinfo{author}{Marchetti, M.},
  \bibinfo{author}{Marenduzzo, D.} \& \bibinfo{author}{Tailleur, J.}
\newblock \bibinfo{title}{{Pattern Formation in Self-Propelled Particles with
  Density-Dependent Motility}}.
\newblock \emph{\bibinfo{journal}{Phys.~Rev.~Lett.}}
  \textbf{\bibinfo{volume}{108}}, \bibinfo{pages}{248101}
  (\bibinfo{year}{2012}).

\bibitem{Solon2013}
\bibinfo{author}{Solon, A.~P.} \& \bibinfo{author}{Tailleur, J.}
\newblock \bibinfo{title}{{Revisiting the Flocking Transition Using Active
  Spins}}.
\newblock \emph{\bibinfo{journal}{Phys.~Rev.~Lett.}}
  \textbf{\bibinfo{volume}{111}}, \bibinfo{pages}{078101}
  (\bibinfo{year}{2013}).

\bibitem{Frey2015}
\bibinfo{author}{Suzuki, R.}, \bibinfo{author}{Weber, C.~A.},
  \bibinfo{author}{Frey, E.} \& \bibinfo{author}{Bausch, A.~R.}
\newblock \bibinfo{title}{{Polar pattern formation in driven filament systems
  requires non-binary particle collisions}}.
\newblock \emph{\bibinfo{journal}{Nat.~Phys.}} 
\textbf{\bibinfo{volume}{11}}, \bibinfo{pages}{839--843}
 (\bibinfo{year}{2015}).

\bibitem{Pearce2015}
\bibinfo{author}{Pearce, D. J.~G.} \& \bibinfo{author}{Turner, M.~S.}
\newblock \bibinfo{title}{{Emergent behavioural phenotypes of swarming models
  revealed by mimicking a frustrated anti-ferromagnet.}}
\newblock \emph{\bibinfo{journal}{J.~R.~Soc.~Interface}} \textbf{\bibinfo{volume}{12}}, \bibinfo{pages}{20150520}
  (\bibinfo{year}{2015}).

\bibitem{Fukui2013}
\bibinfo{author}{Fukui, T.}, \bibinfo{author}{Hatsugai, Y.} \&
  \bibinfo{author}{Suzuki, H.}
\newblock \bibinfo{title}{{Chern Numbers in Discretized Brillouin Zone:
  Efficient Method of Computing (Spin) Hall Conductances}}.
\newblock \emph{\bibinfo{journal}{J. Phys. Soc. Jpn.}}
\textbf{\bibinfo{volume}{74}},
  \bibinfo{pages}{1674--1677} (\bibinfo{year}{2005}).
  
  \bibitem{Brotto}
Brotto, T., Caussin, J.-B., Lauga, E., \& Bartolo, D.
Hydrodynamics of Confined Active Fluids.
\newblock \emph{\bibinfo{journal}{Phys.~Rev.~Lett.}}
\textbf{110}, 038101 (2013).

\bibitem{Landau3}
\bibinfo{author}{Landau, L.~D.}, \bibinfo{author}{Lifshitz, E.~M.},
  \bibinfo{author}{Sykes, J.~B.}, \bibinfo{author}{Bell, J.~S.} \&
  \bibinfo{author}{Rose, M.~E.}
\newblock \emph{\bibinfo{title}{{Quantum Mechanics, Non-Relativistic Theory}}},
  vol.~\bibinfo{volume}{11} (\bibinfo{publisher}{Elsevier Science},
  \bibinfo{year}{1958}).
  
  \bibitem{Vicsek1995}
\bibinfo{author}{Vicsek, T.}, \bibinfo{author}{Czir{\'{o}}k, A.},
  \bibinfo{author}{Ben-Jacob, E.}, \bibinfo{author}{Cohen, I.} \&
  \bibinfo{author}{Shochet, O.}
\newblock \bibinfo{title}{{Novel Type of Phase Transition in a System of
  Self-Driven Particles}}.
\newblock \emph{\bibinfo{journal}{Phys.~Rev.~Lett.}}
  \textbf{\bibinfo{volume}{75}}, \bibinfo{pages}{1226--1229}
  (\bibinfo{year}{1995}).

\bibitem{Levine2000}
\bibinfo{author}{Levine, H.}, \bibinfo{author}{Rappel, W.~J.} \&
  \bibinfo{author}{Cohen, I.}
\newblock \bibinfo{title}{{Self-organization in systems of self-propelled
  particles.}}
\newblock \emph{\bibinfo{journal}{Phys.~Rev.~E}} \textbf{\bibinfo{volume}{63}},
  \bibinfo{pages}{017101} (\bibinfo{year}{2001}).
\end{thebibliography}
\end{document}